\documentclass[aps,prl,preprint,groupedaddress]{revtex4-1}
\usepackage{graphicx}% Include figure files
\usepackage{dcolumn}% Align table columns on decimal point
\usepackage{bm}% bold math

\begin{document}

% Use the \preprint command to place your local institutional report
% number in the upper righthand corner of the title page in preprint mode.
% Multiple \preprint commands are allowed.
% Use the 'preprintnumbers' class option to override journal defaults
% to display numbers if necessary
%\preprint{}

%Title of paper
\title{Effect of dibucaine on phase behavior of ternary liposome}

% repeat the \author .. \affiliation  etc. as needed
% \email, \thanks, \homepage, \altaffiliation all apply to the current
% author. Explanatory text should go in the []'s, actual e-mail
% address or url should go in the {}'s for \email and \homepage.
% Please use the appropriate macro foreach each type of information

% \affiliation command applies to all authors since the last
% \affiliation command. The \affiliation command should follow the
% other information
% \affiliation can be followed by \email, \homepage, \thanks as well.
\author{Kazunari Yoshida}
\email[]{yoshida.k09@nishio-lab.net}
\author{Akito Takashima}
\author{Izumi Nishio}
\email[]{izumi@phys.aoyama.ac.jp}
%\homepage[]{Your web page}
%\thanks{}
\affiliation{College of Science and Engineering, Aoyama Gakuin University, Sagamihara, Kanagawa, 252-5258, Japan}

%Collaboration name if desired (requires use of superscriptaddress
%option in \documentclass). \noaffiliation is required (may also be
%used with the \author command).
%\collaboration can be followed by \email, \homepage, \thanks as well.
%\collaboration{}
%\noaffiliation

\date{\today}

\begin{abstract}
We investigated the effect of Dibucaine hydrochloride (DC$\cdot$HCl), one of the local anesthetics,
on phase behavior of ternary liposome composed of dioleoylphosphatidylcholine (DOPC), dipalmitoylphosphatidylcholine (DPPC), and cholesterol (Chol).
The large DOPC/DPPC/Chol liposome, that is directly observable by optical microscope, is commonly known to be laterally separated into
liquid-ordered (Lo) phase (raft-like domain) and liquid-disordered (Ld) phase under certain conditions and is useful for study of
lipid-raft-like domains as a simple model system.
In order to confirm the effect of DC$\cdot$HCl on a miscibility transition temperature, $T_{\rm c}$, of the ternary liposome,
we observed the liposomes with three concentrations, 0, 0.05, and 0.2~mM, of DC$\cdot$HCl at various temperatures.
In addition, we calculated the angle-averaged two-dimensional autocorrelation (2D-AC) functions in order to quantify the phase behavior.
The results of these observations and calculations revealed that the DC$\cdot$HCl molecules induce the 
reduction of $T_{\rm c}$ of the ternary liposome.
Furthermore, we calculated the circularity of Lo domain in order to confirm the change in the line tension of the Lo/Ld phase boundary
and revealed that the insertion of the DC molecules induces the reduction of line tension.
In terms of the critical phenomena, we conclude that the insertion of the DC molecules induces the reduction of the $T_{\rm c}$ of
the ternary liposome due to reduction of line tension.
This suggests that the DC molecules may disturb function of ion channels via 
affecting the lipid bilayers which surround ion channels.
\end{abstract}

% insert suggested PACS numbers in braces on next line
% insert suggested keywords - APS authors don't need to do this
%\keywords{}

%\maketitle must follow title, authors, abstract, \pacs, and \keywords
\maketitle

% body of paper here - Use proper section commands
% References should be done using the \cite, \ref, and \label commands
\section{Introduction}
Because it is important to understand the physical mechanism of functional expression of anesthetics in the pharmacology,
the effect of anesthetic molecules on biological membrane containing ion channels has been studied for many years.
For instance, effect of chloroform and other general anesthetics on sodium current in a squid giant axon and
interactions of chloroform and/or benzyl alcohol with lipid bilayer have been studied by
using experimental and/or simulation methods\cite{axon,ex_benzyl,ex_chloro,md1,md2}.

Recently, local anesthetics have especially been paid attention and studied.
A lot of studies have reported that local anesthetics affect the sodium current across the cell membrane
\cite{channel,channel2,channel3,channel4,channel5,channel6,channel7}.
In addition, interaction between anesthetics and lipid bilayer
which surrounds the ion channels has also been studied
because it is suggested that the perturbation of the lipids due to invasion of the anesthetics
disturbs the function of ion channels\cite{matsuki1,matsuki2,matsuki3,lc_hcl,fluidity}.
Furthermore, the effect on lipid-raft domains, which are mainly composed of saturated lipids, cholesterol
and membrane proteins such as ion channels, and are dispersed in the cell membranes
(believed to be a kind of phase separation)\cite{raft} should be investigated
in order to clarify the detailed mechanism of local anesthetic function
because the lipid rafts are believed to play important roles in the various cell functions\cite{raft}.
Therefore, the phase-separated liposome with the raft-like domains has been used as a simple model system of cell membrane
for studies of interactions between local anesthetics and lipid bilayers\cite{anes_raft1,anes_raft2}.
However, the detailed physicochemical mechanism of functional expression of local anesthesia is still unclear.

Large phase-separated liposome, that is observable by optical microscope, is useful
for the direct investigations of the raft-like domains.
The multi-component liposome is commonly known to be separated into two or three phases
due to difference of thermodynamical properties of their lipids,
and the phase-separated domains of large liposome are observable by fluorescence microscopy with a dye which is only
incorporated into a certain phase\cite{phase_sep1,phase_sep_v,phase_sep2,phase_sep3,rho2,charge}.
Indeed, large phase-separated liposome has been used for many studies of the lipid membranes,
and the physical properties of the lipid membrane have been revealed\cite{phase_sep1,phase_sep2,phase_sep3,rho2,charge,diffusion,inverse_cone,nucleation}.

In the similar cases to the present study, the large liposome and cell-derived giant plasma membrane vesicle (GPMV) which is similar to the multi-component liposome are also used as a model system.
For example, it has been reported that vitamin E, Triton-X 100, and benzyl alcohol affect the phase morphology of large multi-component liposomes\cite{direct1}.
Gray {\it et al.} have also reported that the general anesthetics including alcohols reduce the 
miscibility transition temperature, $T_{\rm c}$, of GPMVs\cite{direct2}.
However, a direct observation of the influence of
local anesthetics on the phase behavior of large liposome has not been reported.

In the present study, we investigated the influence of dibucaine hydrochloride (DC$\cdot$HCl), one of the commonly used local anesthetics, on phase behavior of liposome composed of dioleoylphosphatidylcholine (DOPC), dipalmitoylphosphatidylcholine (DPPC), and cholesterol (Chol).
The DOPC/DPPC/Chol liposome is one of the famous systems and is known to be laterally separated into liquid-ordered (Lo) and liquid-disordered (Ld) phases under certain conditions\cite{phase_sep_ter}.
The Lo phase, mainly composed of DPPC and Chol, represents the raft-like domain, and the Ld phase is DOPC-rich region\cite{phase_sep_ter}.
We calculated angle-averaged two-dimensional autocorrelation (2D-AC) function of images of liposomes with three concentrations of DC$\cdot$HCl at various temperatures
in order to estimate the change in $T_{\rm c}$ due to the effect of DC.
The results indicate that the DC molecules reduce the $T_{\rm c}$ of ternary liposome.

The 2D-AC and/or angle-averaged 2D-AC functions have been widely used in the various fields
such as biology\cite{ac1} and physics\cite{ac2,ac3}.
In addition, the angle-averaged 2D-AC analysis was performed in order to clarify the correlation length of GPMV surface in the similar study
to this article\cite{ac_gpmv}.
However, to our knowledge, this is the first report of quantitatively demonstrating the change in $T_{\rm c}$ of ternary liposome due to addition of local anesthetics by calculating the angle-averaged 2D-AC function of membrane domain patterns.

\section{Materials and methods}
\subsection{Materials}
\begin{figure}[htbp]
	\begin{center}
		\includegraphics[width=65mm]{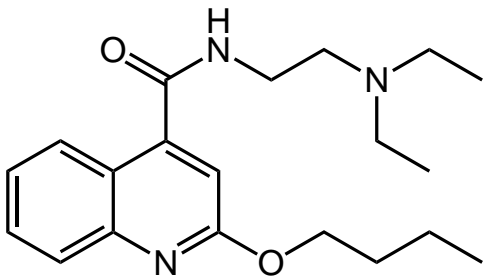}
		\caption{Chemical structure of DC.}
		\label{fig:dibucaine}
	\end{center}
\end{figure}

Dibucaine hydrochloride (DC$\cdot$HCl), DOPC (chain melting temperature, $T_{\rm m} \approx -18~^\circ\mathrm{C}$)\cite{transition},
DPPC ($T_{\rm m} \approx 41~^\circ\mathrm{C}$)\cite{transition}, Chol and chloroform were purchased from Wako Pure Chemical Industries, Ltd. (Japan). The chemical structure of DC is described in Figure~\ref{fig:dibucaine}. Methanol was purchased from Showa Chemical Industry Co., Ltd. (Japan).
Rhodamine B 1,2-dihexadecanoyl-sn-glycero-3-phosphoethanolamine
(rhodamine~DHPE, $\lambda_{\rm ex}=560~{\rm nm},~\lambda_{\rm em}=580~{\rm nm}$), 
fluorescent phospholipid, was obtained from Invitrogen (U.S.A.). Ultra-pure water was
obtained using a WT101UV AUTOPURE (Yamato Scientific Co., Ltd., Japan).

\subsection{Preparation of Ternary Liposome}
We prepared liposomes using the natural swelling methods\cite{hydration,yoshida},
The lipids of DOPC/DPPC/Chol = 50/25/25~(mol\%) with 0.5~mol\% of rhodamine~DHPE were dissolved in mixture solvent of chloroform
and methanol with a volume ratio of 2:1.
Then, we removed the organic solvents through an air flow and placing the sample in an aspirated desiccator for more than 8 hours
in order to make a dry lipid film.
Finally, liposomes formed through hydration of the lipid film with the ultra-pure water at 37~$^\circ\mathrm{C}$ for more than 24 hours.
The lipid concentration was 0.2~mM.

\subsection{Observation}
40~$\mu$L of liposome suspension and 40~$\mu$L of DC$\cdot$HCl aqueous solution were gently mixed,
and the solution was incubated at room temperature (RT, 21.7~$\pm$~0.4 $^\circ\mathrm{C}$) for more than 10 minutes
in order to wait for sufficient dispersion of DC$\cdot$HCl molecules into the lipid membrane.
In this study, we used 0, 0.1, and 0.4~mM of DC$\cdot$HCl aqueous solutions. 
The final concentration of lipids was 0.1~mM with 0, 0.05, or 0.2~mM of DC$\cdot$HCl.
Then, 5~$\mu$L of the mixed solution was placed between two cover glasses and was sealed with vacuum grease.
The sample was placed on a copper plate, and the temperature of the plate was controlled using a peltier device (deviation $<0.1~^\circ\mathrm{C}$) with monitoring the plate temperature using a thermistor probe placed in the copper plate (under the sample).
We put a thin layer of thermal grease between the sample cell and the copper plate
in order to keep the high thermal conduction from the plate to the sample.
The sample placed on the plate was incubated for more than 10 minutes in order to archive the thermal equilibrium state before observation
and was observed using a fluorescence microscope BX40 (Olympus, Japan) through a objective lens of 40$\times$ at various temperature.
We recorded the microscopic images using a digital camera DP73 (Olympus, Japan).
Excitation-light irradiation ($\lambda_{\rm max}=546$~nm) was applied using a Hg lamp through a WIG filter set (Olympus, Japan).

\subsection{Analysis of microscopic images}
We calculated the angle-averaged 2D-AC using a macro program of the ImageJ\cite{imagej},
First, fluorescence microscopic images of liposome surface were cut into 64 $\times$ 64 pixels and were converted into gray scale (16~bit).
Then, we calculated two-dimensional fast Fourier transform (2D-FFT) using brightnesses of each pixel of images and inverse transform of their power spectrum
in order to derive 2D-AC function.
Further, the 2D-AC function was averaged for all angles.
Finally, we calculated the FFT of $G(r)/G(0)$ curves, where $G(r)$ is a angle-averaged 2D-AC as a function of
radial position, $r$, in the real space.
Typical results of the analysis of one- and two-phase images are described
in the Appendix A.

\subsection{Calculation of Lo-domain circularity}
The circularity was also calculated using ImageJ\cite{imagej}.
We transformed the microscopic images to binary images in order to obtain the phase boundary of Lo/Ld.
The circularity, dimensionless parameter, of the Lo domains was defined as
\begin{equation}
{\rm (circularity)}= \frac{4\pi A}{l^2},
\end{equation}
where the $\pi$ is the ratio of the circumference of a circle to its diameter, $A$ is a area of Lo domain, and $l$ is a perimeter of Lo domain.
The circularity of 1 corresponds to perfect circle.
We calculated the circularity in order to estimate the change in line tension of phase boundary in the case of $T=20~^\circ {\rm C}$.

\section{Results and discussion}
\begin{figure}[htbp]
	\begin{center}
		\includegraphics[width=130mm]{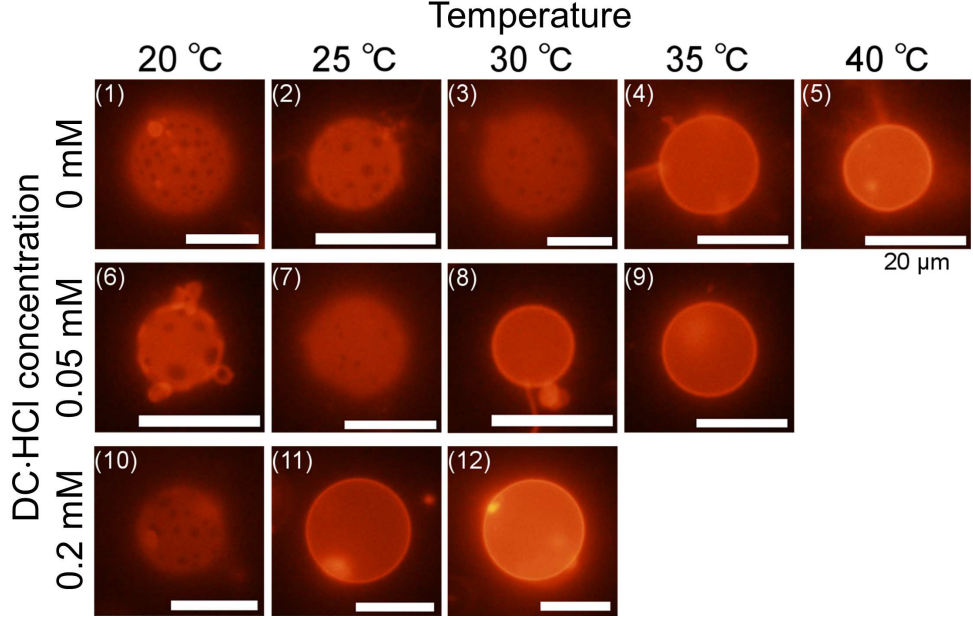}
		\caption{Typical microscopic images of DOPC/DPPC/Chol liposomes
				with 0, 0.05, and 0.2~mM of DC$\cdot$HCl at 20, 25, 30, 35, and 40~$^\circ\mathrm{C}$.
				Scale bars represent 20~$\mu$m.}
		\label{fig:images}
	\end{center}
\end{figure}

We observed the DOPC/DPPC/Chol liposomes with 0, 0.05, and 0.2~mM of DC$\cdot$HCl at $T=20$, 25, 30, 35, and $40~^\circ\mathrm{C}$
in order to confirm the influence of DC$\cdot$HCl on phase behavior of the ternary liposomes.
Figure~\ref{fig:images} shows the typical microscopic images at each condition.
Since rhodamine DHPE is mainly localized in Ld phase of the phase-separated systems\cite{rho1,rho2},
the dark region corresponds to DPPC- and Chol-rich domain (Lo, raft-like domain) while the bright region corresponds to DOPC-rich domain (Ld).
Image~1 described in the Figure~\ref{fig:images} shows the phase-separated liposome with scattered Lo domains.
The number of phase-separated liposome was reduced with increasing temperature (images~1-5).
In the case of 0~mM and at $35~^\circ {\rm C}$,
we observed many one-phase liposomes.
There is no phase-separated liposome at $40~^\circ {\rm C}$, corresponding to image~5.
Next, we also observed the ternary liposomes at various temperature in the case of 0.05~mM DC$\cdot$HCl.
We observed the phase separation in the most of liposomes with 0.05~mM DC$\cdot$HCl at $20~^\circ {\rm C}$, corresponding to image~6.
There are several phase-separated liposomes with smaller Lo domains as shown in the
image~7 and a few liposomes with large domains as shown in Figure~\ref{fig:large_sep} described in Appendix B at 25~$^\circ {\rm C}$.
We observed a few phase-separated liposomes with smaller Lo domains at $30~^\circ {\rm C}$ (image~8)
, and most of liposomes dose not have the Lo domains at 35~$^\circ {\rm C}$ (image~9).
Further, in the case of 0.2~mM, the number of phase-separated liposomes was also reduced with increasing temperature.
At 20~$^\circ {\rm C}$, phase-separated liposomes are majority, however, there are several liposomes which have 
unclear phase boundaries.
At 25~$^\circ {\rm C}$ and above, most of liposomes have one phase without raft-like domains, corresponding to images~11 and 12.
We especially focus on the images at 25~$^\circ {\rm C}$ with the three DC$\cdot$HCl concentrations,
corresponding to images 2, 7, and 11.
The Lo domains in the liposome surfaces became smaller with increasing DC$\cdot$HCl concentration.
These results indicate that DC molecule have an ability to change the phase behavior of DOPC/DPPC/Chol liposomes.
It is considered that the effect of HCl is ignorable because the lipids used in this study is electrically neutral (not charged lipids)\cite{rho2,charge}.

\begin{figure}[htbp]
	\begin{center}
		\includegraphics[width=130mm]{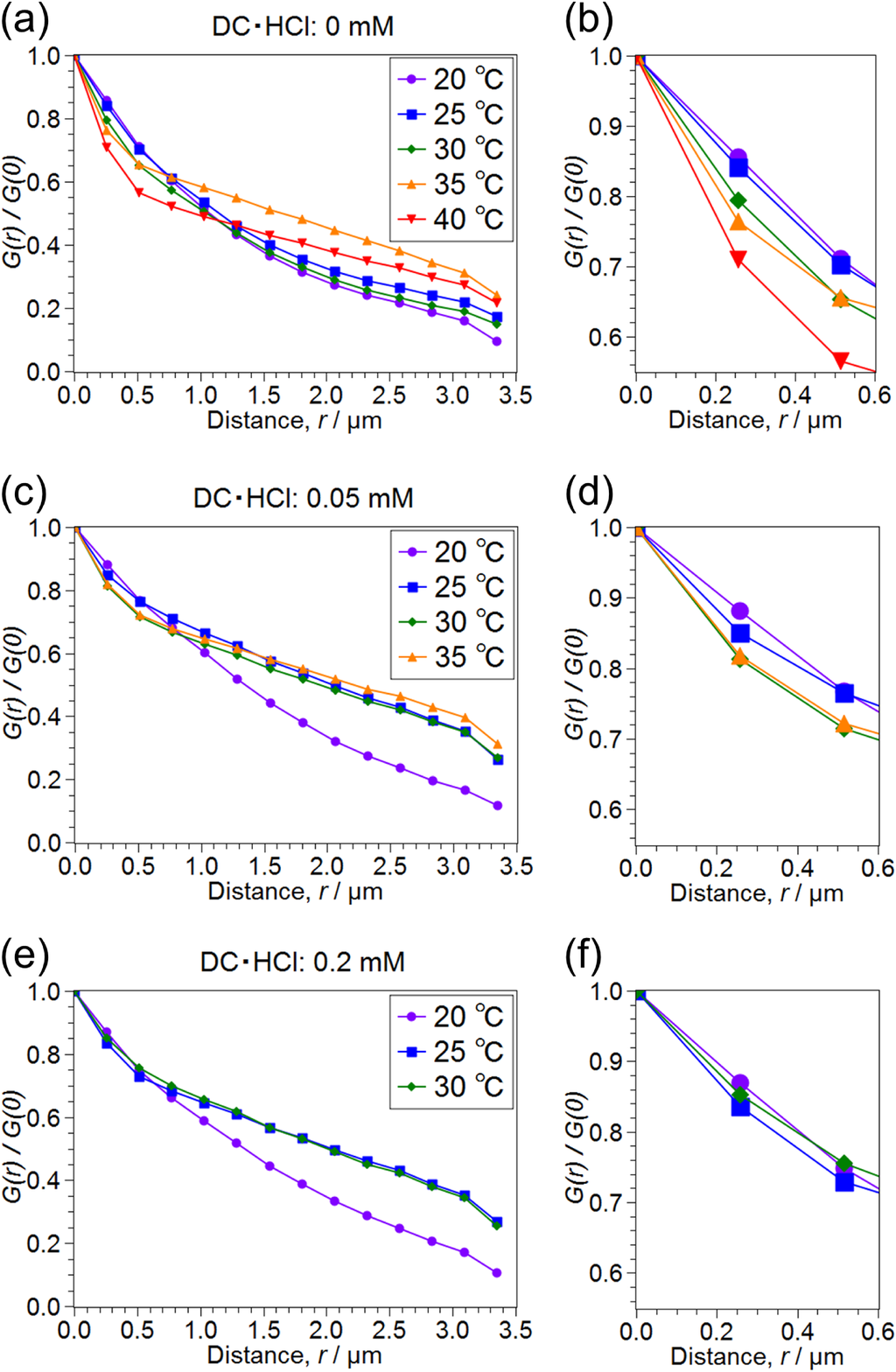}
		\caption{Mean values of the angle-averaged 2D-AC functions ($n \geq 10$ at each condition, error bars not shown).
				(a, c, and e) The angle-averaged 2D-AC functions of ternary liposomes
				with 0, 0.05, and 0.2~mM of DC$\cdot$HCl, respectively.
				(b, d, and f) Enlarged view of the short distance region of the angle-averaged 2D-AC function,
				corresponding to (a, c, and e), respectively.}
		\label{fig:2d_ac1}
	\end{center}
\end{figure}

In order to quantify the phase behavior, 
we calculated the 2D-AC functions of domain pattern in the ternary liposomes.
Figures~\ref{fig:2d_ac1}(a, c, and e) show mean values of the normalized angle-averaged 2D-AC, $G(r) / G(0)$, 
as a function of radial position, $r$, at 0, 0.05, and 0.2~mM, respectively,
and Figures~\ref{fig:2d_ac1}(b, d, and f) show the enlarged view of the short distance region of the angle-averaged 2D-AC functions,
corresponding to Figures~\ref{fig:2d_ac1}(a, c, and e), respectively.
The numbers of the observed liposomes are not less than 10 at each condition.
Value of the short distance region (especially $r \approx 0.25~\mu$m) of the $G(r)/G(0)$ 
decreases with increasing temperature.
It is considered that the correlation of the short distance region dependents on the ratio of the phase-separated liposomes.
The correlation of the phase-separated liposome tends to have higher value due to Lo domains than that of the one-phase liposome
as shown in Figure~S1.
The results indicate that the ratio of the phase-separated liposomes is reduced with increasing temperature.
In the case of the 0.2~mM, dispersion of the values at $r \approx 0.25~\mu$m is smaller than that of two other concentrations
because the differences of ratio of phase-separated liposomes between three temperatures seem to be smaller than that of 0 and 0.05~mM.

\begin{figure}[htbp]
	\begin{center}
		\includegraphics[width=130mm]{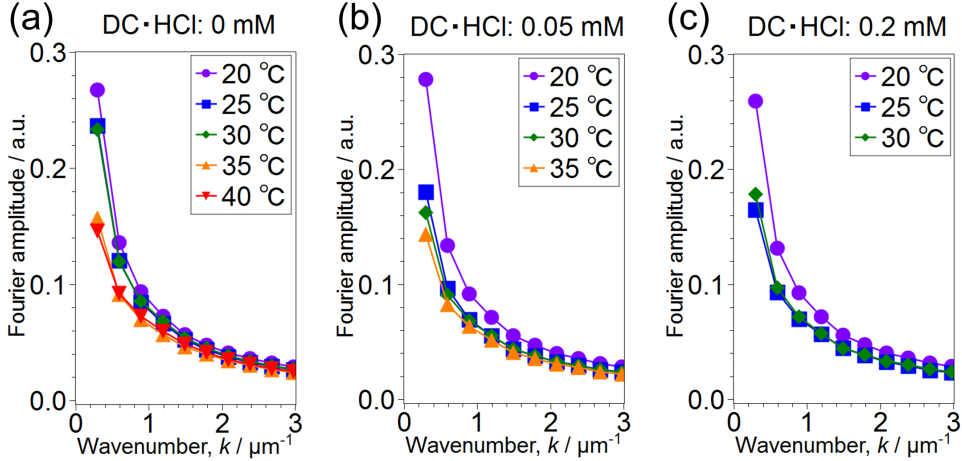}
		\caption{(a, b, and c) Mean values of the Fourier amplitudes of the $G(r)/G(0)$ curves
				as a function of wavenumber, $k$, at 0, 0.05, and 0.2~mM, respectively
				($n \geq 10$ at each condition, error bars not shown).}
		\label{fig:fft_amp}
	\end{center}
\end{figure}

Since we need further informations in order to quantitatively confirm the effect of DC$\cdot$HCl molecules on miscibility transition temperature, $T_{\rm c}$,
FFT of the $G(r)/G(0)$ curves of individual liposomes was calculated.
Figures~\ref{fig:fft_amp}(a, b, and c) show the Fourier spectrum of the $G(r)/G(0)$ curves at 0, 0.05, and 0.2~mM, respectively ($n \geq 10$).
Fourier amplitude of phase-separated liposome at wavenumber $k \approx 0.3$ and $0.6~\mu$m$^{-1}$ tends to have larger value
due to Lo domains than that of one-phase one as shown in Figure~S1,
where the wavenumber $k \approx 0.3$ and $0.6~\mu$m$^{-1}$ correspond to
 wavelength $\lambda \approx 3.3$ and $1.7~\mu$m, respectively.
Therefore, the values of the Fourier amplitudes depend on the rate of phase-separated liposomes.
The amplitudes at $k \approx 0.3$ and $0.6~\mu$m$^{-1}$ are reduced with increasing temperature at all concentrations.
In addition, largest distance point of Fourier amplitudes corresponds from 30 to 35~$^\circ {\rm C}$ at 0~mM (Figure~\ref{fig:fft_amp}(a)),
while the points correspond from 20 to 25~$^\circ {\rm C}$ at 0.05 and 0.2~mM (Figures~\ref{fig:fft_amp}(b and c)).
This indicates that the DC$\cdot$HCl molecules reduce the $T_{\rm c}$ because it is considered that 
such largest distance point roughly corresponds to $T_{\rm c}$.

\begin{figure}[htbp]
	\begin{center}
		\includegraphics[width=65mm]{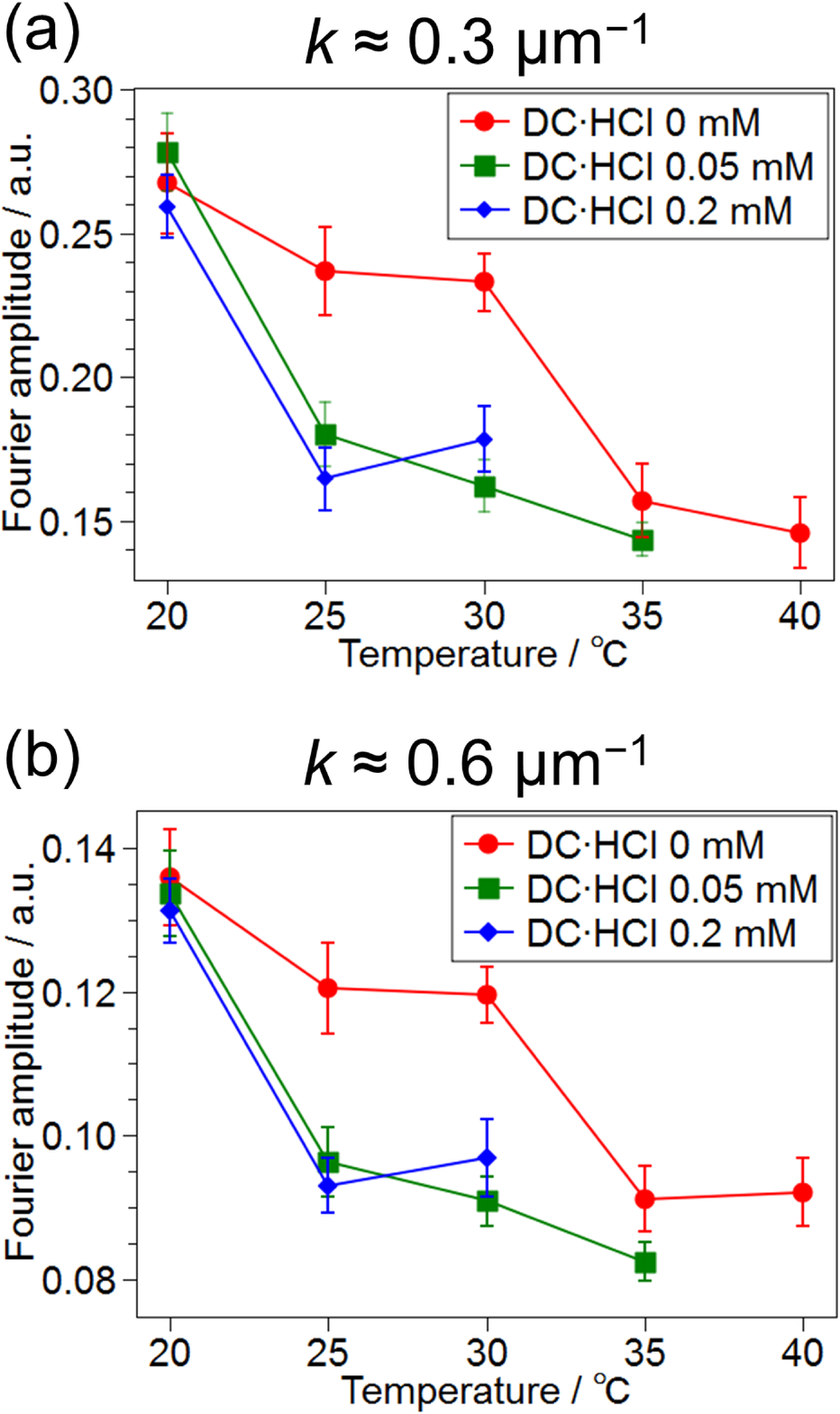}
		\caption{(a and b) The Fourier amplitude versus temperature at $k\approx 0.3$ and 0.6~$\mu$m$^{-1}$, respectively.
					Error bars represent standard errors ($n \geq 10$ at each condition).}
		\label{fig:amp_temp}
	\end{center}
\end{figure}

Furthermore, relation between Fourier amplitude and temperature was shown in Figure~\ref{fig:amp_temp}. 
The result indicates that the $T_{\rm c}$ is reduced with increasing DC$\cdot$HCl concentration.
The $T_{\rm c}$ downwardly shifts more than 5~$^\circ {\rm C}$ from 0~mM to 0.05 and 0.2~mM.
The difference of Fourier amplitude between 0.05 and 0.2~mM at 25~$^\circ {\rm C}$ is 
larger at $k\approx 0.3~\mu$m$^{-1}$ than at $k\approx 0.6~\mu$m$^{-1}$.
This indicates that the amount of the liposomes with Lo domains corresponded to $k \approx 0.3~\mu$m$^{-1}$ is larger at 0.05~mM
than at 0.2~mM,
and that the amount of the liposomes with Lo domains corresponded to $k \approx 0.6~\mu$m$^{-1}$ at 0.05~mM is similar to at 0.2~mM.
In other words, the ratio of liposomes having the large domains is higher at 0.05~mM than at 0.2~mM
at 25~$^\circ {\rm C}$.
This suggests that the difference of the DC$\cdot$HCl concentration influences the phase behavior.
To our knowledge, this is the first report of quantitatively showing the change in $T_{\rm c}$ of DOPC/DPPC/Chol liposome 
due to effect of local anesthetic molecule by 2D-AC analysis.

\begin{figure}[htbp]
	\begin{center}
		\includegraphics[width=65mm]{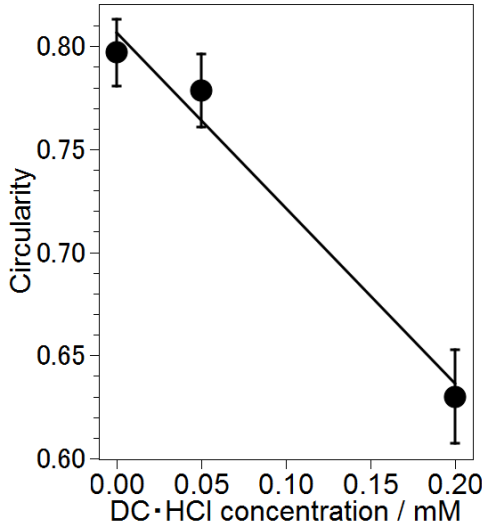}
		\caption{Circularity of the Lo domains versus DC$\cdot$HCl concentration at 20~$^\circ {\rm C}$ ($<T_{\rm c}$).
				Error bars represent standard errors ($n \geq 27$ Lo domains at each concentration).}
		\label{fig:circ}
	\end{center}
\end{figure}

In this study, we demonstrated the change in $T_{\rm c}$ of DOPC/DPPC/Chol liposome due to effect of DC$\cdot$HCl molecules.
We discuss the cause of change in $T_{\rm c}$ by calculating a circularity of the Lo domains.
Figure~\ref{fig:circ} shows the circularity, $4\pi A/l^2$, at $20~^\circ {\rm C}$.
The circularity is reduced with increasing the DC$\cdot$HCl concentration.
This suggests that the line tension of the Lo/Ld phase boundary is reduced with the DC$\cdot$HCl concentration
because the domain circularity is related to line tension of phase boundary\cite{afm}.
In other words, the DC molecule induces the reduction of line tension of phase boundary.
It was previously shown that the DC molecules has an ability to invade the lipid bilayers\cite{nmr},
which led us to consider this possibility in the case of DOPC/DPPC/Chol systems, and the effect of HCl is ignorable as described above.
In addition, line-tension change induced by incorporation of other molecules was reported by several groups\cite{tension1,tension2}. 
Therefore, it is considered that the insertion of the DC molecules into bilayers induces the reduction of the line tension.

Previously, line tension of the phase boundary was expressed as
\begin{equation}
\sigma \propto (T_c - T)^\nu,
\label{eq:tension}
\end{equation}
where the $\sigma$ is a line tension of the Lo/Ld phase boundary,
and the $\nu$ is a critical exponent\cite{phase_sep1,critical}.
In terms of this equation, the miscibility transition temperature, $T_{\rm c}$, is reduced with the decrease in the line tension, $\sigma$,
when the temperature of the system, $T$, is constant.
Therefore, it is considered that the insertion of the DC molecules into bilayers induces the reduction of the $T_{\rm c}$ (Figure~\ref{fig:amp_temp})
through the reduction of the line tension. 

A similar studies to the present work suggested that the general anesthetics disturb the phase separation of liposome and GPMV (reduction of $T_{\rm c}$)
by using the neutron diffraction and/or X-ray diffraction methods\cite{inhalation1,inhalation2} and the direct observation\cite{direct2}.
It is believed that the behavior in these previous studies is also induced by insertion of the anesthetic molecules to bilayers.
Actually, Hamada {\it et al.} have reported that photoisomerization of the azobenzene derivative reversibly switches the phase pattern
due to change in lateral line tension\cite{photo}.

\section{Conclusion}
In summary, we have investigated the influence of DC$\cdot$HCl molecules, one of the local anesthetics, on phase behavior of DOPC/DPPC/Chol $=50/25/25$ (mol\%) liposome.
In order to confirm the effect of DC$\cdot$HCl molecules, the ternary liposomes with 0, 0.05, and 0.2~mM of DC$\cdot$HCl have been observed
at various temperatures.
As a result, we have clarified the DC molecules has an ability to change the miscibility transition temperature, $T_{\rm c}$ of 
the ternary liposome by calculating the angle-averaged 2D-AC function.
Furthermore, we calculated a circularity of the Lo domains at $20~^\circ {\rm C}$ in order to estimate the change in line tension of
the Lo/Ld phase boundary
and revealed that insertion of the DC molecules into bilayers induces the reduction of the $T_{\rm c}$ via decrease in line tension.
This suggests that the DC molecules disturb the function of the ion channels (anesthetic function of DC) through affecting the lipid bilayers which surround the ion channels.
Since these physicochemical findings hint at the mechanism of the anesthetic function,
this study may play an important role in the pharmacology.

\section{Acknowledgements}
The authors thank Dr.~Yasuhiro Fujii (Ritsumeikan Univ.), Prof.~Tsutomu Hamada (JAIST),
Dr.~Rina Kagawa (M.D., Univ. Tokyo), and Prof.~Miho Yanagisawa (Tokyo Univ. Agr. Tech.)
for their significant advises.

\clearpage
\appendix
\section{Appendix A: Two-dimensional aoutocorrelation (2D-AC) analysis}
\begin{figure}[htbp]
	\begin{center}
		\includegraphics[width=117mm]{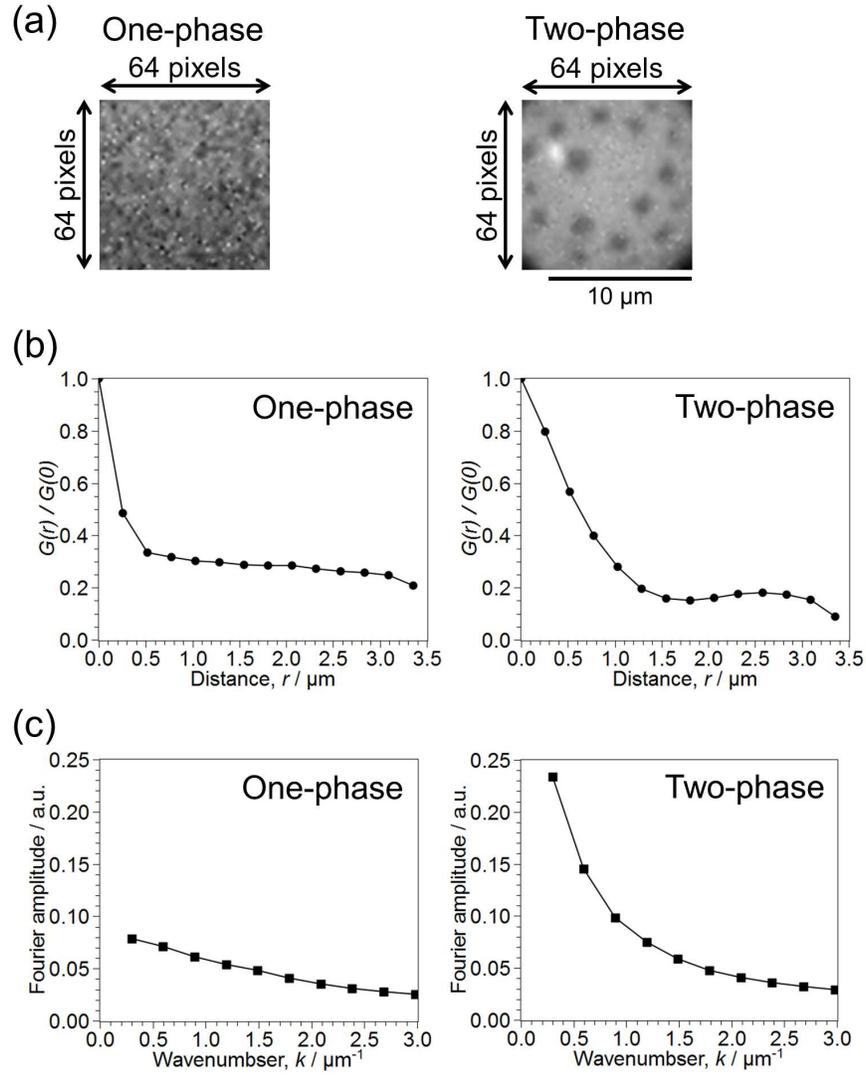}
		\caption{(a)~Typical fluorescence microscopic images of one- and two-phase ternary liposomes. 
					(b)~Angle-averaged 2D-AC functions, $G(r)/G(0)$, of each domain pattern, corresponding to (a).
					(c)~Fourier spectrum of each domain pattern, corresponding to (a and b).}
		\label{fig:scheme}
	\end{center}
\end{figure}

\clearpage
\section{Appendix B: Microscopic image of other domain pattern}
\begin{figure}[htbp]
	\begin{center}
		\includegraphics[width=55mm]{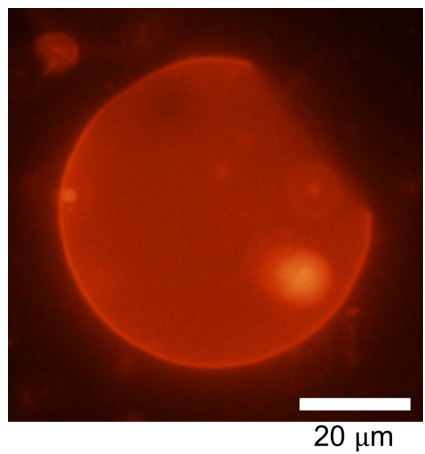}
		\caption{Example of ternary liposome with 0.05~mM of DC$\cdot$HCl at 25~$^\circ\mathrm{C}$.
					Scale bar represents 20~$\mu$m.}
		\label{fig:large_sep}
	\end{center}
\end{figure}

\bibliography{ref_dib}

\end{document}